\documentclass[fleqn,usenatbib,useAMS]{mnras}
\usepackage{graphicx,xfrac,amsmath,longtable,color}
\usepackage{float}
\usepackage[caption = false]{subfig}
\def\la{\mathrel{\hbox{\rlap{\hbox{\lower4pt\hbox{$\sim$}}}\hbox{$<$}}}}
\def\ga{\mathrel{\hbox{\rlap{\hbox{\lower4pt\hbox{$\sim$}}}\hbox{$>$}}}}

\def\kms{km~s$^{-1}$}

\def\dm15{{$\Delta$}$m_{15}$}

\def\v10{$V_{10}$(Si~II)}

\def\W575{$W(5750)$}
\def\W610{$W(6100)$}
\def\6100{the 6100~\AA\ absorption}

\def\m{M$_\odot$}
\def\msun{M$_\odot$}

\def\Lsun{L$_\odot$}
\def\rsun{R$_\odot$}

\def\gcm3{g~cm$^{-3}$}
\def\cm2g{cm$^{2}$~g$^{-1}$}

\def\CaII7291{[Ca~{\sc II}] $\lambda\lambda$7291,7323}

\def\OI6300{[O~{\sc I}] $\lambda\lambda$6300,6364}

\def\apj{ApJ}
\def\apjl{ApJL}
\def\apjs{ApJS}
\def\aj{AJ}
\def\nat{Nature}
\def\mnras{MNRAS}
\def\aap{A\&A}

\def\pasj{PASJ}

\usepackage[T1]{fontenc}
\usepackage{ae,aecompl}

\title[Betelgeuse Rotation]{The Betelgeuse Project: Constraints from Rotation}

\author[J. C. Wheeler et al.]{
J. Craig Wheeler,$^{1}$\thanks{Contact e-mail: wheel@astro.as.utexas.edu} S. Nance,$^{1}$ M. Diaz,$^{1}$
S. G. Smith,$^{1}$ J. Hickey,$^{1}$ L. Zhou,$^{2}$
\newauthor M. Koutoulaki,$^{3}$ J. M. Sullivan,$^{1}$ J. M. Fowler$^{4}$\\
$^{1}$Department of Astronomy, University of Texas at Austin, Austin, TX, USA\\
$^{2}$Department of Physics, Tsinghua University\\
$^{3}$Dublin Institute for Advanced Studies\\
$^{4}$Tufts University
}

\begin{document}

\pubyear{2016}

\label{firstpage}
\pagerange{\pageref{firstpage}--\pageref{lastpage}}
\maketitle

\begin{abstract}

In order to constrain the evolutionary state of the red supergiant Betelgeuse 
($\alpha$ Orionis) we have produced a suite of models with ZAMS masses from 
15 to 25 \msun\ in intervals of 1 \msun\ including the effects of rotation. The 
models were computed with the stellar evolutionary code {\it MESA}. For 
non--rotating models we find results that are similar to other work. It is 
somewhat difficult to find models that agree within 1$\sigma$ of the observed 
values of R, T$_{eff}$ and L, but modestly easy within 3$\sigma$ uncertainty. 
Incorporating the nominal observed rotational velocity, $\sim$15~\kms, yields 
significantly different, and challenging, constraints. This velocity constraint 
is only matched when the models first approach the base of the red supergiant 
branch (RSB), having crossed the Hertzsprung gap, but not yet having ascended 
the RSB and most violate even generous error bars on R, T$_{eff}$ and L. Models 
at the tip of the RSB typically rotate at only $\sim$0.1~\kms, independent of any 
reasonable choice of initial rotation. We discuss the possible uncertainties in 
our modeling and the observations, including the distance to Betelgeuse, the 
rotation velocity, and model parameters. We summarize various options to account 
for the rotational velocity and suggest that one possibility is that Betelgeuse 
merged with a companion star of about 1~\msun\ as it ascended the RSB, in the process 
producing the ring structure observed at about 7' away. A past coalescence would 
complicate attempts to understand the evolutionary history and future of Betelgeuse.

\end{abstract}



\section{Introduction}

Betelguese ($\alpha$ Orionis) is a massive red supergiant that is destined to 
explode as a Type IIP supernova and leave behind a neutron star. One of us (JCW) has 
long been obsessed with the uncertainty in the evolutionary state of Betelgeuse and 
has sought means to reduce that uncertainty. This effort, in various guises, has been 
informally deemed {\sl The Betelgeuse Project.} An evolving team of undergraduates 
has participated in this project. Here we report on results on Betelgeuse itself, 
especially the constraint of its rotational state. Non--rotating models aimed at 
reproducing the observed aspects of Betelgeuse are given by \citet{2013EAS....60...17M} 
and \citet{2016ApJ...819....7D}. \citeauthor{2016ApJ...819....7D} found the rather 
surprising result that the favored mass of Betelgeuse was $\sim19$~\msun, somewhat 
larger than popular estimates.

The ultimate question of determining when Betelgeuse will explode depends on 
knowing the current evolutionary state. A typical basic model of 20 \msun\ 
begins core helium burning as the model crosses the Hertzsprung gap. The 
model is still in core helium burning when it first hits the tip of the red supergiant
branch (RSB) at a luminosity of $L \approx 10^5$ \Lsun. The model then forms a 
semi-convective hydrogen-burning shell and retreats down the RSB to 
$L = 10^{4.83}$ \Lsun, still substantially brighter than the minimum luminosity
at the base of the RSB, $L = 10^{4.55}$ \Lsun. The model ends core helium 
burning with a luminosity of $L = 10^{4.95}$ \Lsun. Core carbon burning is
initiated at $L = 10^{5.1}$ \Lsun, only about 2000 years before core collapse. 
Betelguese is most probably in some phase of core helium burning.

\section{Computations}

We evolved a grid of models from the Zero Age Main Sequence (ZAMS) to near the 
onset of core collapse using the stellar evolution code Modules for Experiments in 
Stellar Astrophysics ({\it MESA}; \citealt{2011ApJS..192....3P, 2013ApJS..208....4P,
2015ApJS..220...15P}). We computed models of solar metallicity with ZAMS masses 
from 15 to 25 \msun\ at intervals of 1 \msun\ primarily using {\it MESA} versions 6208 
and 7624.

For one suite, the models were non--rotating. For another suite with the same 
ZAMS masses we assumed an initial rotation of 200~\kms, corresponding to about 
25\% of the Keplerian velocity on the ZAMS. As we will discuss below, our results 
do not depend sensitively on the particular choice of this initial rotational 
velocity; a lower value would give lower final velocities and a significantly 
larger initial velocity would be unphysical. \citet{2016ApJ...819....7D} did a 
careful exploration of the sensitivity of the choice of convective and overshoot 
schemes. Because our principal goal was to explore the effect of rotation, we 
chose only the default prescriptions in {\it MESA}: Schwarzschild convection 
and an overshoot parameter of $\alpha = 0.2$. For the rotating models, we again 
chose  {\it MESA} default values of mechanisms of angular momentum transport 
and mixing. We did not include magnetic effects \citep{2015ApJ...799...85W}. 
We employed the ``Dutch" mass--loss prescriptions with $\eta = 0.8$. We used 
nuclear reaction network {\sl approx21}. The inlist we employed is available 
upon request from the authors. For each ZAMS mass, the models were computed 
to the onset of core collapse. In practice, we have presented data at the
end of core carbon burning. At those stages, there is little change in the 
outer, observable, characteristics. In the following discussion we refer to 
specific models by their ZAMS mass.

\section{Results}
\label{results}

We compared the results of our models to the observational constraints on 
Betelgeuse as given by \citet{2016ApJ...819....7D}: log$L/L_{\odot} = 5.10\pm0.22$; 
$R/R_{\odot} = 887\pm203$; $T_{eff} = 3500\pm200$K. \citet{2016ApJ...819....7D}
considered a range of possible uncertainties, including limb darkening and overshoot,
but the uncertainties in $L$ and $R$ are dominated by the uncertainty in distance 
$D = 197\pm45$ pc \citep{2008AJ....135.1430H}, with $R \propto D$ and $L \propto D^2$. 
We recognize that the errors in the distance estimate are likely to be distributed 
asymmetrically and subject to the Lutz--Kelker--Hanson (LKH) bias that will tend to
favor larger distances and hence larger radii and luminosities \citep{2007AJ....133.1810B},
but for perspective we have also examined the constraints on models considering 
naive $3\sigma$ errors, simply three times the $1\sigma$ errors. For $T_{eff}$, the 
$3\sigma$ range is $\pm 600$~K and for $R/R_{\odot}$, $\pm 608$. For $L$, we find 
$L/L_{\odot} = 1.3^{+0.7}_{-0.5}\times10^5$ and $L/L_{\odot} = 1.3^{+2.4}_{-1.2}\times10^5$ 
for the $1\sigma$ and $3\sigma$ ranges, respectively. In terms of logarithms, these 
values and ranges are log~$T_{eff} = 3.54^{+0.03 (+0.07)}_{-0.02 (-0.08)}$, 
log~$(R/R_{\odot}) = 2.95^{+0.09 (+0.22)}_{-0.11 (-0.51)}$, and log~$(L/L_{\odot}) = 
5.1^{+0.19 (+0.47)}_{-0.21 (-0.99)}$, where the $3\sigma$ limits are given in 
parentheses. In addition, we have employed the constraint that the observed 
rotational velocity of Betelgeuse is $v~sin~i \approx 5$~\kms\ with an inclination 
of $i  \approx 20^o$ \citep{1987ApJ...317L..85D,1996ApJ...463L..29G,1998AJ....116.2501U,
2009A&A...504..115K}. These 
data imply an equatorial rotational velocity of $\sim 15$~\kms.
The uncertainty in this quantity is itself uncertain.

In principle, the surface gravity of Betelgeuse provides another independent constraint
on the ratio $R/M$. \cite{1984ApJ...284..223L} adopted log~$g = 0.0\pm0.3$. 
\cite{2000ApJ...545..454L} obtained log~$g = -0.5$ and \cite{2011ASPC..451..117N}
determined $R/M = 82^{+13}_{-12}$(\rsun/\msun). From their best--fit models,
\citet{2016ApJ...819....7D} obtained $R/M = 40$(\rsun/\msun) and log~$g = -0.05$
for their Eggleton--based code and log~$g = -0.10$ with {\it MESA}.

\subsection{Non--rotating Models}
\label{nonRot}

\begin{figure}
\center
\includegraphics[width=3 in, angle=0]{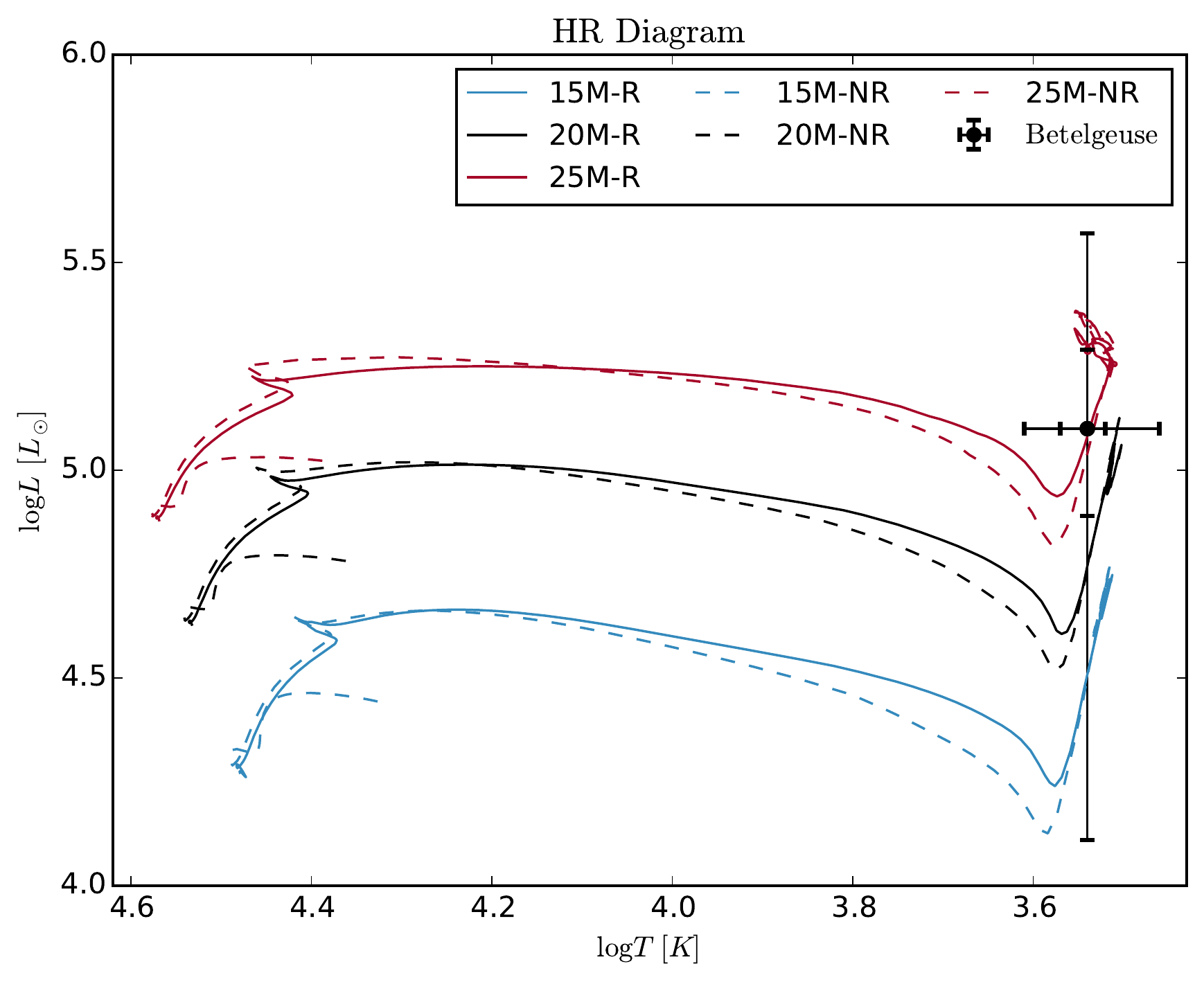}
\caption
{The evolutionary tracks of the models of 15, 20, and 25 \msun. Dashed
lines correspond to non--rotating models. Solid lines correspond
to models that began with an equatorial velocity of 200~\kms\ on the
ZAMS. The observed position of Betelgeuse and the adopted $1\sigma$ 
and $3\sigma$ error bars are also shown.
\label{HRD}}
\end{figure}


\begin{table}[h!]
\hspace{-.1in}
\begin{tabular}{p{3.0in}}

\includegraphics[width = 3.0in, angle=0]{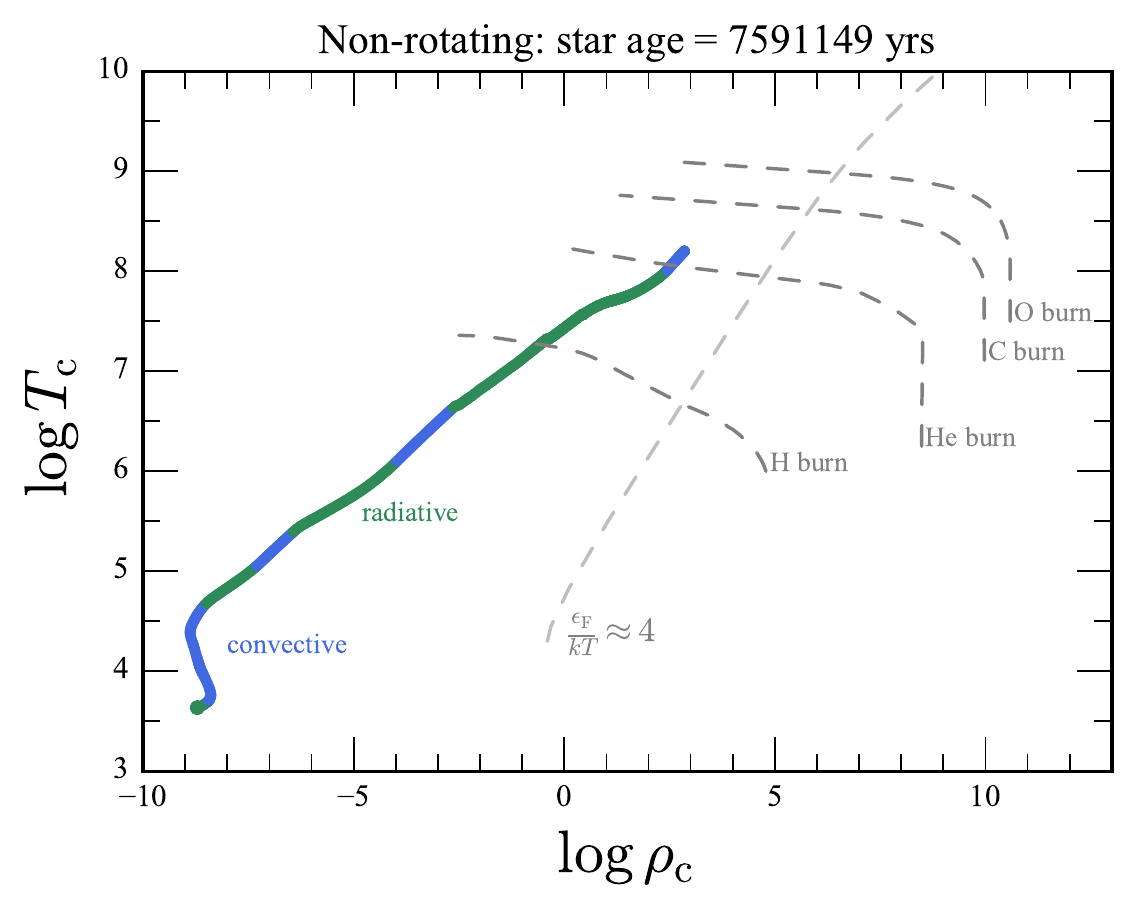}\\
\includegraphics[width = 3.0in, angle=0]{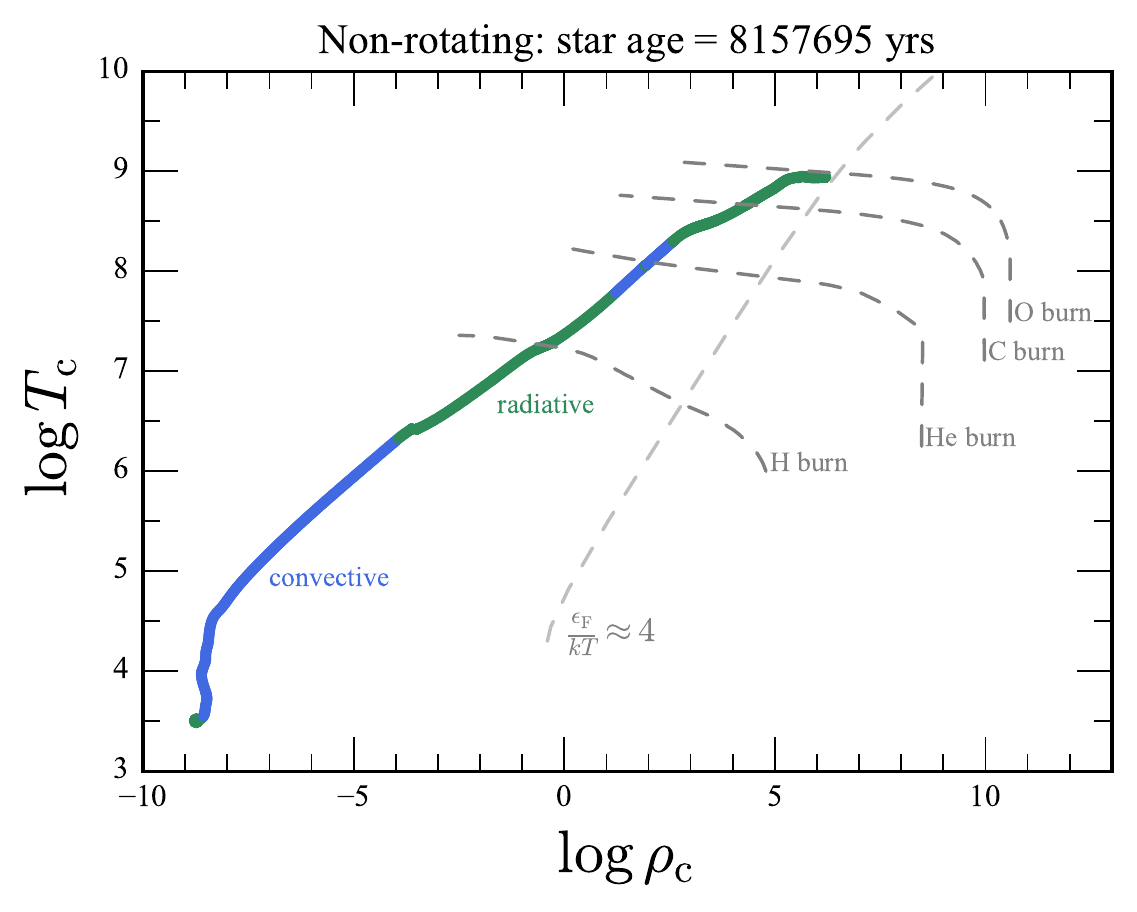}\\
{\bf Figure 2} Structure of the non--rotating 20~\m\ model in the temperature/density 
plane in core helium burning near the luminosity minimum (panel a) and near 
the end of carbon burning (panel b). The blue regions are convective, the 
green regions, radiative. The dashed line corresponding to $\epsilon_F/(kT)\approx4$
represents the non-degenerate/degenerate boundary.
\\
\label{trho}

\end{tabular}
\end{table}


\setcounter{figure}{2}

Figure \ref{HRD} shows representative evolutionary tracks in the Hertszprung--Russell 
diagram for the 15, 20 and 25 \msun\ models, both rotating and non--rotating and 
the corresponding position of Betelgeuse. Figure \ref{trho} shows the interior structure 
of the non--rotating 20~\msun\ model at the point of minimum luminosity at the base
of the RSB and at the end of core carbon burning, defined as the central mass 
fraction of $^{12}$C being less than $\sim 10^{-4}$. The structure of the models 
in both panels of Figure \ref{trho} is characteristic of all the models, roughly 
independent of mass. An exception is the right panel which does not show
an inner region of convective carbon burning in this particular snapshot. This
inner convective region is common in models and is shown in panel b of Figure 
\ref{trhorot}.

Betelgeuse can be brought into agreement with either the minimum luminosity 
point at the base of the RSB or at the tip of the RSB with a judicious choice 
of distance within the errors. We note that our models do not yield any blue 
loops, consistent with the results of \citet{2016ApJ...819....7D}, but that 
the minimum luminosity of our models is somewhat dimmer than those of 
\citeauthor{2016ApJ...819....7D}. We find that the rotating models give 
somewhat brighter luminosities at the luminosity minimum at the base of the 
RSB in a manner consistent with the results of \citet{2013EAS....60...17M}. 
Convective overshoot has rather little effect on the evolutionary track but 
can also increase the luminosity at the minimum \citep{2016ApJ...819....7D}. 
The surface abundances of Betelgeuse are enhanced in nitrogen, depleted in 
carbon and display a low ratio of $^{12}C/^{13}C$, all signs that Betelgeuse 
has undergone the first dredge--up and hence that it must be on the ascending 
RSB \citep{2016ApJ...819....7D}. As discussed in \S\ref{Rot}, this constraint 
is in conflict with the results of basic rotating models. 


\begin{table}[h!]
\hspace{-.1in}
\begin{tabular}{p{3.0in}}
\includegraphics[width = 3.0in, angle=0]{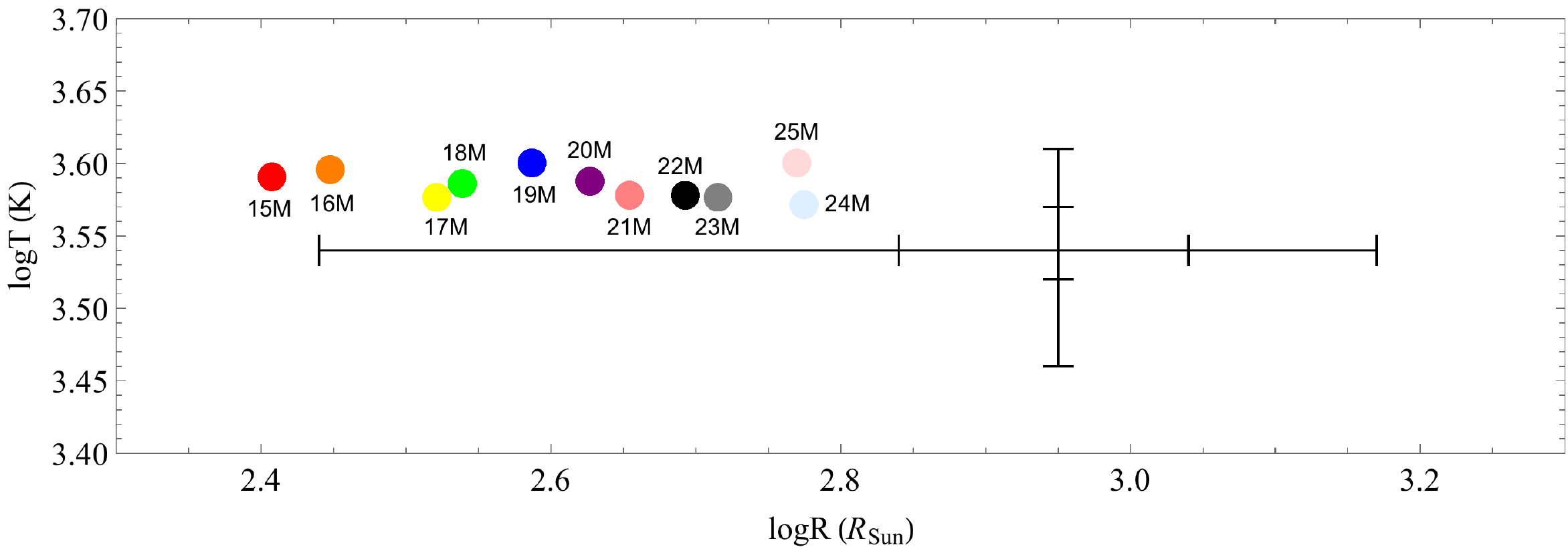}\\
\includegraphics[width = 3.0in, angle=0]{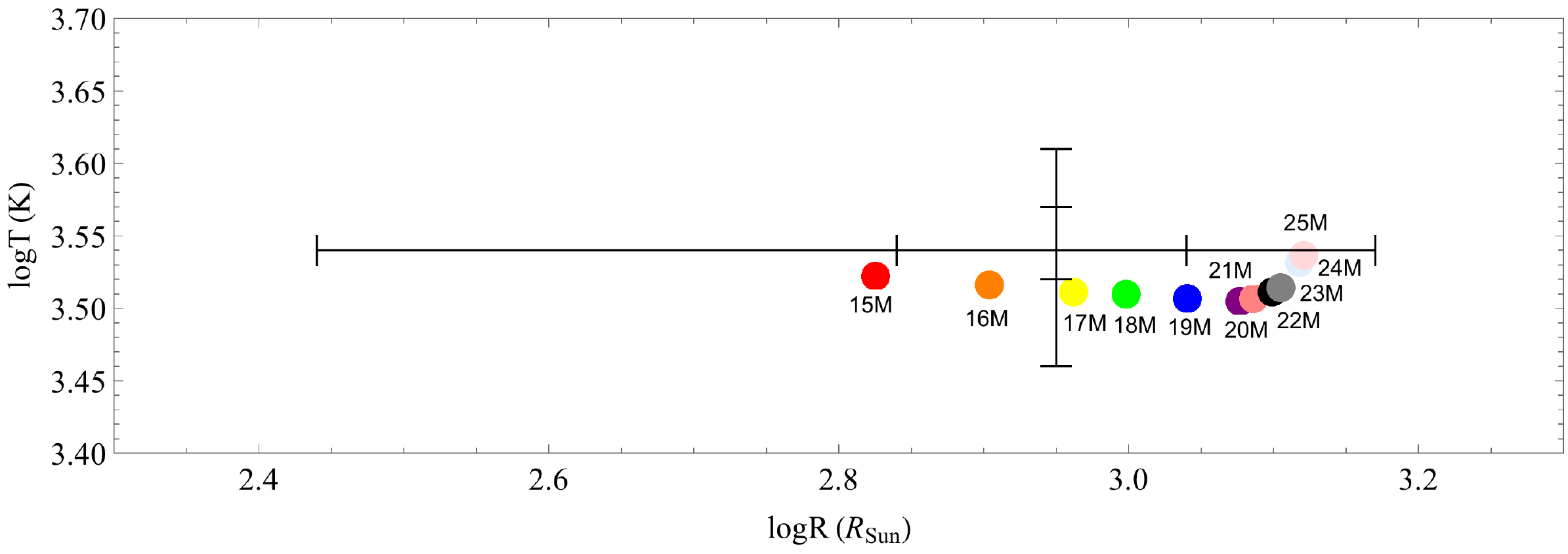}\\
\includegraphics[width = 3.0in, angle=0]{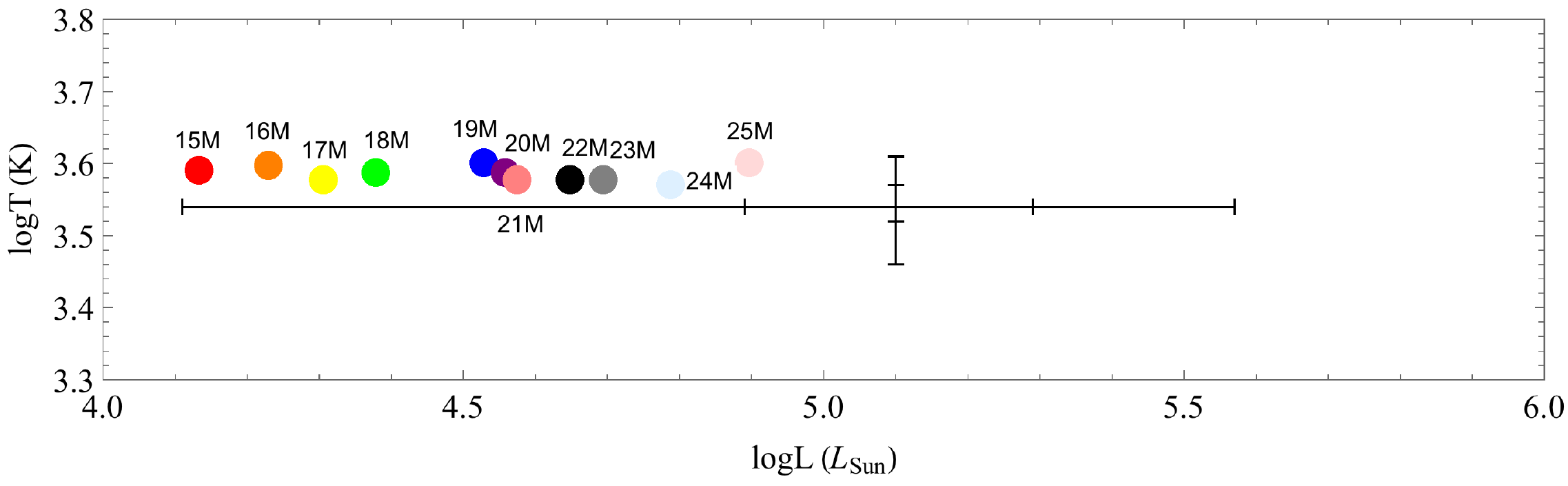}\\
\includegraphics[width = 3.0in, angle=0]{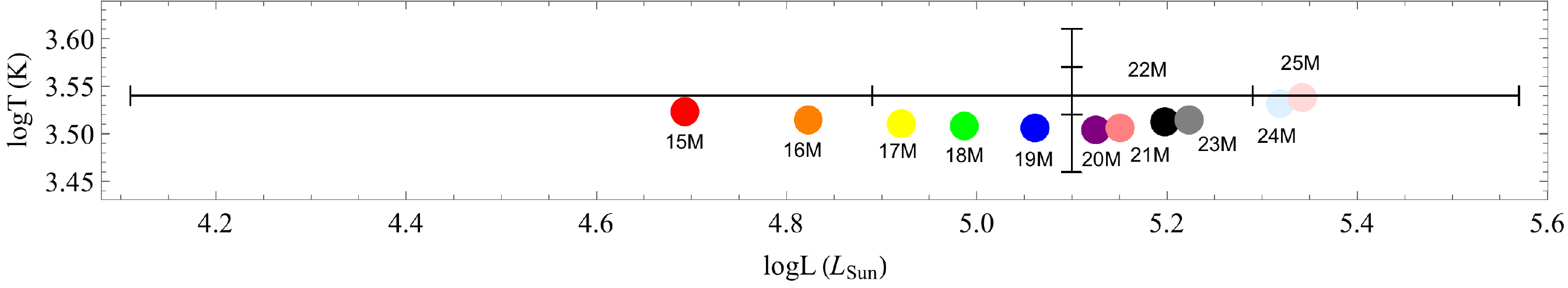}\\
{\bf Figure 3} The distribution of $T_{eff}$ and $R$ is given at the base of the RSB 
where the luminosity is a local minimum for the non--rotating models (panel a) 
and at the end of carbon burning (panel b). The distribution  of $T_{eff}$ and 
$L$ is given at the base of the RSB (panel c) and at the end of carbon burning 
(panel d). The point of minimum luminosity is near where the rotating models 
reproduce the observed rotation rate (\S\ref{Rot}). The observed values for 
Betelgeuse are given with tic marks at the $1\sigma$ and $3\sigma$ limits.
Note the different temperature scales in each panel.
\\
\label{RLTnonrot}
\end{tabular}
\end{table}

\setcounter{figure}{3}

As we will show, the rotating models yield a velocity consistent with the 
observations near the luminosity minimum at the base of the RSB. While this 
constraint does not apply to the non--rotating models, we illustrate conditions 
near the luminosity minimum. Figure \ref{RLTnonrot} (panel a) gives the distribution 
of $T_{eff}$ and $R$ at the point of minimum luminosity for the non--rotating 
models. All the models tend to be too small and hot to agree with the observations 
within $1\sigma$, but they mostly agree within $3\sigma$. A judicious adjustment 
of the uncertain distance might then bring general agreement for any of the 
individual models. Figure \ref{RLTnonrot} (panel b) gives the distribution of 
$T_{eff}$ and $R$ at the end of core carbon burning for the non--rotating 
models. Models 15 to 19 match the radius to within $1\sigma$; all models agree 
within $3\sigma$. The temperature basically matches well with models 16 to 23 
being slightly too cool, and the radii of individual models would agree with an 
appropriate choice of distance. 

Similar results pertain to the distribution of $T_{eff}$ and $L$ at the luminosity
minimum as shown in Figure \ref{RLTnonrot} (panel c). The models tend to be 
too dim to agree with the observations within $1\sigma$. They all 
essentially agree within $3\sigma$. A judicious adjustment of the distance to
smaller values than we have assumed would again bring agreement
for a given model with model 25 being the best fit to the luminosity. 
Figure \ref{RLTnonrot} (panel d) gives the distribution of $T_{eff}$ and $L$ at the 
end of core carbon burning for the non--rotating models. Models 17 to 23 match 
the luminosity to within $1\sigma$; all models agree within $3\sigma$.
The luminosity of individual models would agree with an appropriate choice of distance. 
 
The models that formally most closely match the observations of $T_{eff}$, $R$, 
and $L$ in our suite of non--rotating models is model 24 at the minimum 
luminosity and models 17, 18, and 19 at the end of carbon burning.

For our 20~\msun\ model, the non--rotating models give log~$g = +0.02$
at the luminosity minimum and log~$g = -0.3$ during carbon burning. The
former is nominally too large, and the latter in better agreement with the
determination of \cite{2000ApJ...545..454L}, but it is difficult to rule the
former out, given the level of uncertainty.

\subsection{Rotating Models}
\label{Rot}

Figure \ref{trhorot} gives the structure of the 20~\msun\ model when the
equatorial velocity is near 15~\kms\ and the corresponding rotating model
near the end of carbon burning. Note that the structure in core helium 
burning is very similar to that given in the non--rotating model at the base 
of the RSB in Figure \ref{trho}. The right hand panel shows the characteristic 
structure at that late phase, including the region of vigorous, off--center, carbon 
burning that occurs in many of the other models.  

Figure \ref{RLTrot} gives the distribution of $T_{eff}$ and $R$ when the surface 
velocity is $v_{rot}\sim 15$~\kms\ for the rotating models near the base of the 
RSB (panel a). Figure \ref{RLTrot} shows most of the models are too hot to
agree even within the $3\sigma$ limit. All the models basically agree in radius 
within the 3$\sigma$ limit. The model of 21~\msun\ comes closest to agreeing with 
both $T_{eff}$ and $R$ within 3$\sigma$. Figure \ref{RLTrot} also gives the 
distribution of $T_{eff}$ and $R$ at the end of core carbon burning for the 
rotating models (panel b). Models 15 to 20 match the radius to within $1\sigma$; 
all models agree within $3\sigma$. The temperature basically matches well, with 
the lower--mass models falling slightly low. The radius would agree for all 
models with a judicious choice of distance. 


\begin{table}[h!]
\hspace{-.1in}
\begin{tabular}{p{3.0in}}

\includegraphics[width = 3.0in, angle=0]{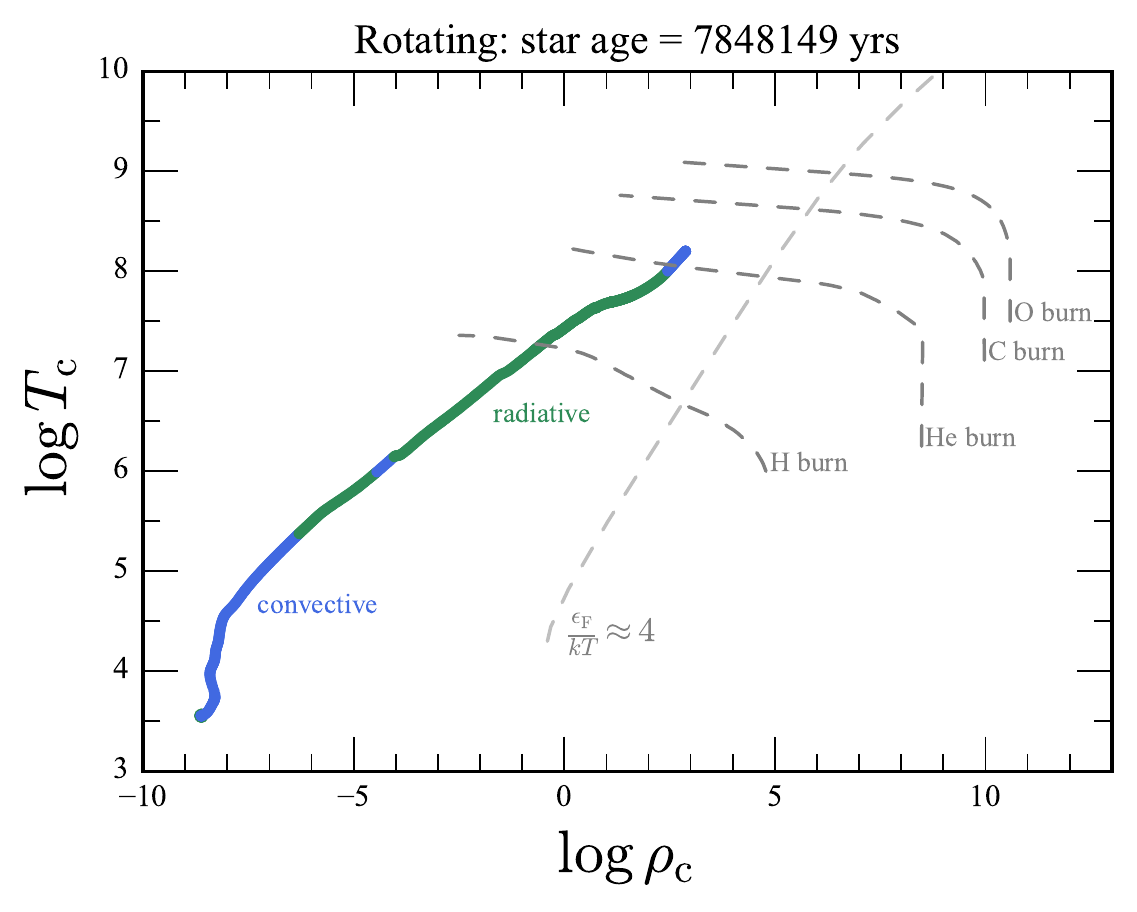}\\
\includegraphics[width = 3.0in, angle=0]{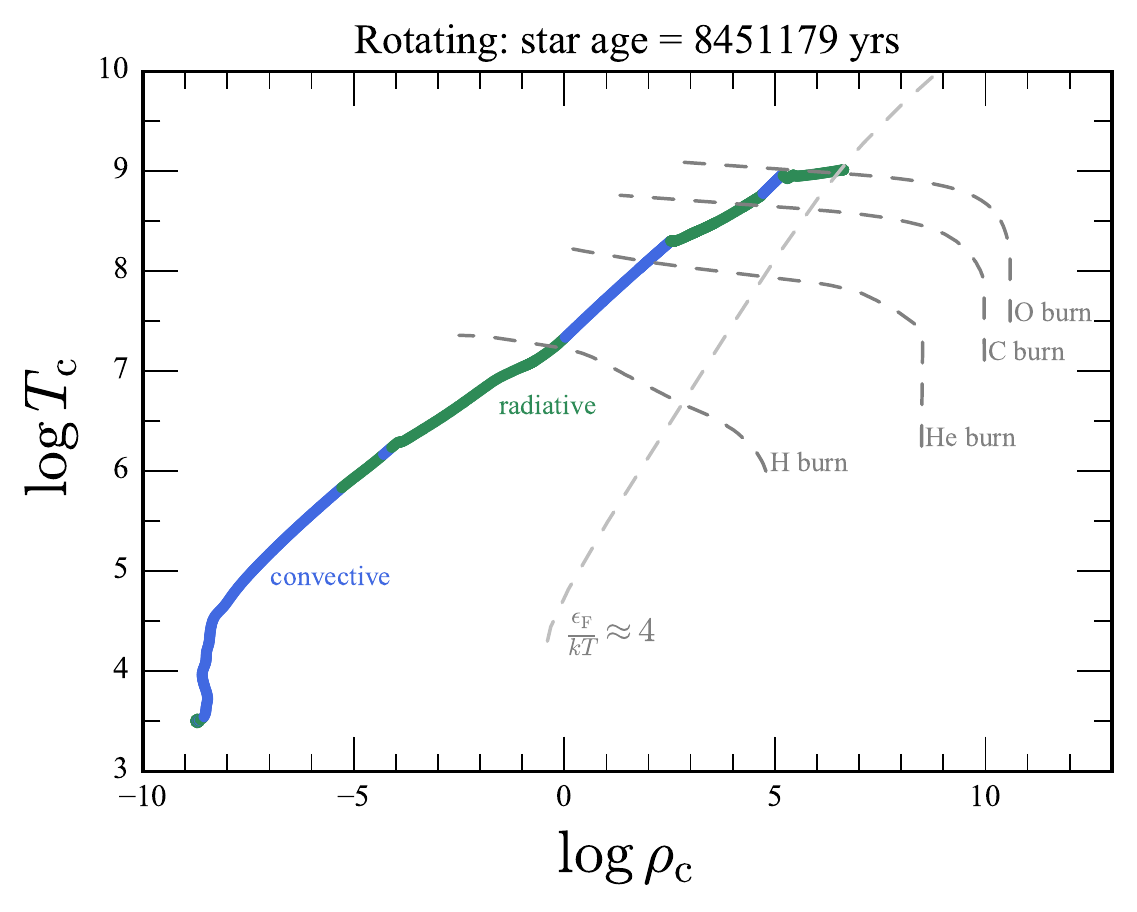}\\
{\bf Figure 4} Structure of the rotating 20~\m\ model in the temperature/density 
plane in core helium burning when the equatorial velocity is 15~\kms\ (panel a) 
and near the end of carbon burning (panel b). The blue regions are convective, the 
green regions, radiative. The dashed line corresponding to $\epsilon_F/(kT)\approx4$
represents the non-degenerate/degenerate boundary.
\\
\label{trhorot}

\end{tabular}
\end{table}


\setcounter{figure}{4} 


\begin{table}[h!]
\hspace{-.1in}
\begin{tabular}{p{3.0in}}

\includegraphics[width = 3.0in, angle=0]{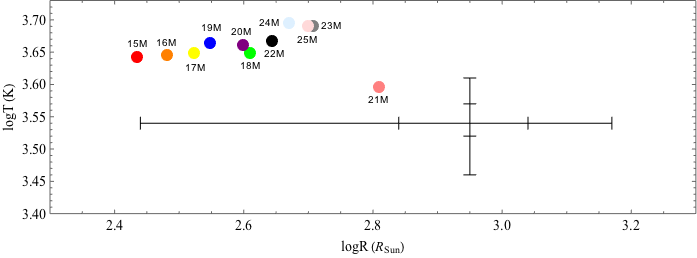}\\
\includegraphics[width = 3.0in, angle=0]{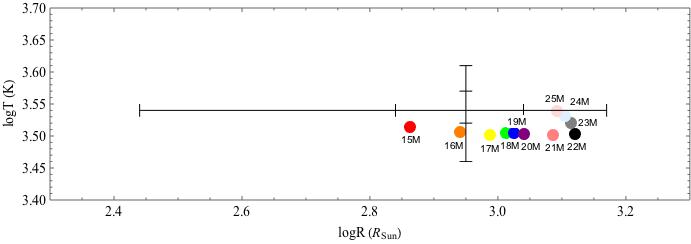}\\
\includegraphics[width = 3.0in, angle=0]{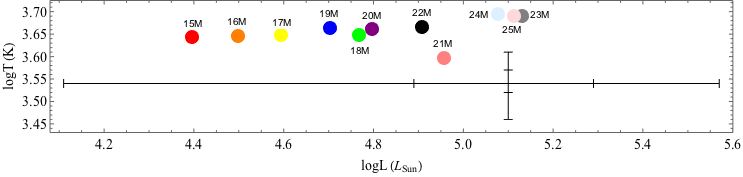}\\
\includegraphics[width = 3.0in, angle=0]{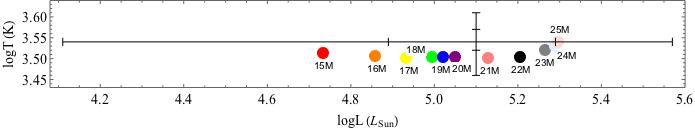}\\
{\bf Figure 5} The distribution of $T_{eff}$ and $R$ is given for our rotating models
at the point at the base of the RSB when the surface velocity is $\sim 15$~\kms\  
(panel a) and at the end of carbon burning (panel b). The distribution of $T_{eff}$ 
and $L$ is given for our rotating models at the point when the surface velocity 
is $\sim 15$~\kms\ (panel c) and at the end of carbon burning (panel d) when 
the surface velocity is $\sim0.1$~\kms. The observed values for Betelgeuse are 
given with tic marks at the $1\sigma$ and $3\sigma$ limits.
\\
\label{RLTrot}

\end{tabular}
\end{table}

\setcounter{figure}{5} 

Similar results pertain to the distribution of $T_{eff}$ and $L$ when the surface 
velocity is $v_{rot}\sim 15$~\kms\ for the rotating models near the base of the 
RSB as shown in Figure \ref{RLTrot} (panel c). The luminosity of Models 22 to 25
agree with $L$ within $1\sigma$, the lower--mass models within $3\sigma$. Model 21 
agrees with the luminosity within $1\sigma$, but again in temperature only at 
$3\sigma$. An appropriate adjustment of the distance could bring general agreement
in $L$ but not in $T_{eff}$. Figure \ref{RLTrot} also gives the distribution of 
$T_{eff}$ and $L$ at the end of core carbon burning for the rotating models 
(panel d). Models 17 to 23 match the luminosity to within $1\sigma$; all models 
agree within $3\sigma$. The temperature basically matches well, and the luminosity 
would agree with a judicious choice of distance. 

The models that formally most closely match the observations of $T_{eff}$, $R$, 
and $L$ in our suite of rotating models are model 21 at the point where
$v_{rot}\sim 15$~\kms\ and models 16 to 20 at the end of carbon burning, ignoring
the velocity constraint. While in some sense a special case, Model 21 with 
nominally the observed rotation of Betelgeuse cannot be formally ruled out by 
the data. This model could be accommodated if Betelgeuse were somewhat hotter 
than the nominal value of 3500 K and the distance were somewhat closer than the 
nominal 197 pc. More models could be accommodated if the rotational velocity 
were taken to be $v_{rot}\sim 5$~\kms.

For our 20~\msun\ model, the rotating models give log~$g = +0.42$
at the luminosity minimum and log~$g = -0.48$ during carbon burning. The
former is substantially too large and the latter in close agreement with the
determination of \cite{2000ApJ...545..454L}.

\begin{figure}
\center
\includegraphics[width=3 in, angle=0]{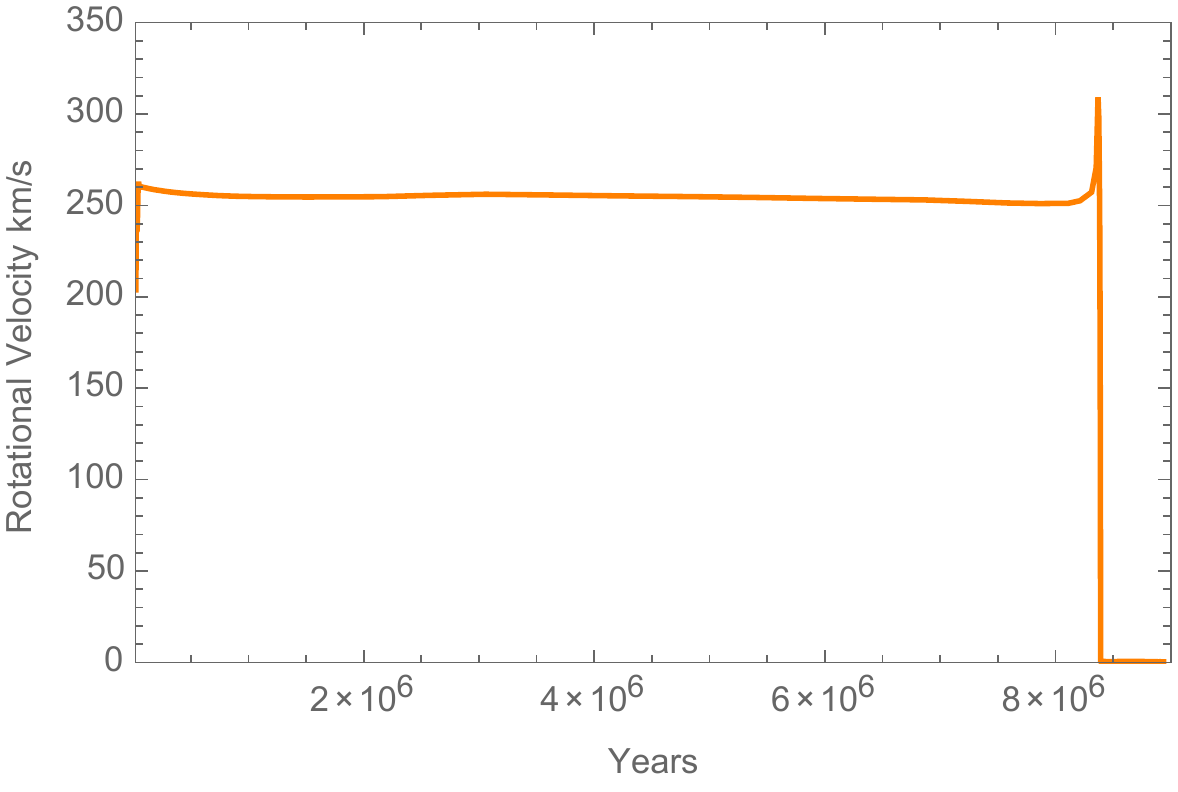}
\caption
{The average surface velocity as a function of time for the model of 20~\msun. 
\label{surf_avg_v_rot-vs-age-20TOT}}
\end{figure}

The rotating models yield the opportunity to examine the rotational state and 
compare to the observed value for Betelgeuse, $v_{rot} \sim 15$~\kms. Figure 
\ref{surf_avg_v_rot-vs-age-20TOT} gives the evolution of the average surface 
velocity for the model with ZAMS mass of 20~\msun. The models relax a little 
as they settle onto the ZAMS and typically begin there with a velocity somewhat 
greater than 200~\kms. The velocity has a small spike, to $v_{rot} \sim300$~\kms, 
due to the contraction at the end of core hydrogen burning, then begins a rapid 
plummet as the models cross the Hertzsprung gap and proceed up the RSB. Models 
at the tip of the RSB that nominally reproduce the observations of $T_{eff}$,
$R$, and $L$ for Betelgeuse typically rotate at only $\sim$0.1~\kms, independent 
of any reasonable choice of initial rotation. While the uncertainty in the 
observed velocity is difficult to assess, this value at the tip of the RSB is 
far below any credible value. The rotation of Betelgeuse represents a dilemma.

\begin{figure}
\center
\includegraphics[width=3 in, angle=0]{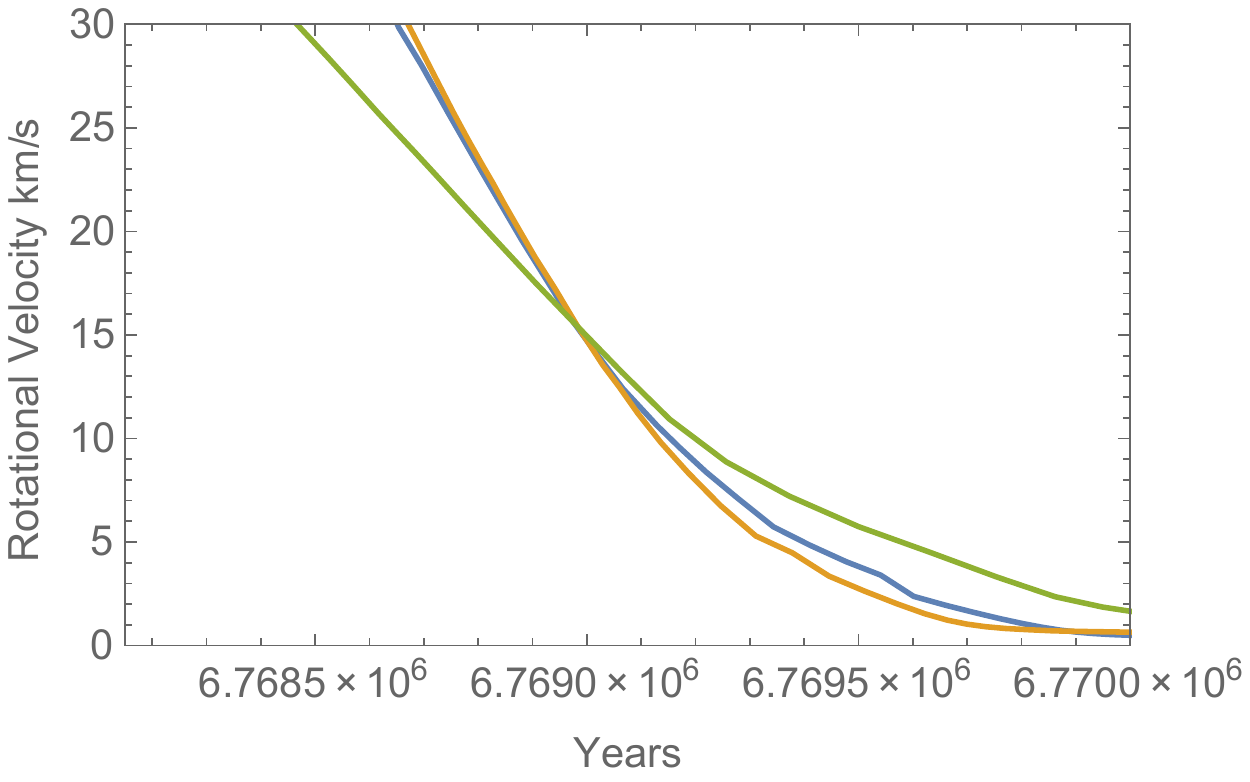}
\caption
{The average surface velocity as a function of time for the model of 
15~\msun\ (green), 20~\msun\ (orange) and 25~\msun\ (blue) near the base 
of the RSB when $v_{rot}\sim 15$~\kms. The timescale given on the abscissa 
is for that of model 25. The other curves have been shifted in absolute time 
so that they align at 15~\kms, but the differential times with respect to 
the epoch of 15~\kms\ are representative. 
\label{velocity_blowup}}
\end{figure}

Figure \ref{RLTrot} gives the distribution of results of $R$, $T_{eff}$ and $L$ 
at the point where the models give a surface rotation velocity close to 15~\kms. 
These models correspond to a very peculiar, special condition. They represent 
the evolutionary stage after the models have crossed the Hertzsprung gap and 
are at very nearly the point of minimum luminosity before the sharp rise up the 
RSB. Figure \ref{velocity_blowup} gives a blow--up of the velocity evolution of 
model 20 from Figure \ref{surf_avg_v_rot-vs-age-20TOT} and also for models 15 
and 25 during the brief epochs when the models gives $v_{rot} \sim 15$~\kms. 
The velocity plunges through the observed range and on down to $v_{rot} 
\sim0.1$~\kms\ as the model rises along the RSB. The 15~\msun, 20~\msun, and 
25~\msun\ models pass through the interval 20~\kms\ to 10~\kms\ in about 400 y. 

These results deepen the dilemma represented by the observed rotation of
Betelgeuse. The only models that formally fit the observed data on $L$, 
$R$, $T_{eff}$ and $v$ are required to sit at a very special, short--lived point 
in the evolution. At the end of the main sequence, these stars cross the 
Hertzsprung gap in the hydrogen shell--burning phase. The shell tends to 
sit at a node at fixed radius while the helium core contracts toward helium 
ignition and the envelope expands. Some of the internal energy is expended 
in inflating the envelope and the luminosity drops as the radius expands 
and the temperature declines. The minimum in luminosity occurs as the envelope 
makes the transition from being radiative to having a deep convective 
structure. From that point, the star nearly reprises its proto--star phase 
in reverse, climbing up the Hyashi track at nearly constant temperature. 
For a typical model, the time to cross the Hertzsprung gap is $\sim 3\times10^4$ y. 
The time to evolve from the point of minimum luminosity and maximum post--ZAMS 
rotation to carbon burning is $\sim 6\times10^5$ y. The interval in which 
the models are predicted to have conditions similar to Betelgeuse and $v_{rot}$ 
between 1 and 50~\kms\ is $\sim$ 1000 years. The probability that Betelgeuse 
happens to be in this special state of transition is low. In addition, these 
models give an excessively large gravity.

We also note that the hypothesis that Betelgeuse is near the point of minimum 
luminosity would require it to be in a stage of expanding radius. This could be 
inconsistent with the observation by \citet{2009ApJ...697L.127T} that the radius 
of Betelgeuse systematically decreased by 15\% from 1993 to 2009, about 1\% 
per year (but see \citealt{2013EAS....60..121O}). In the vicinity of the luminosity 
minimum at the base of the giant branch, our 20~\msun\ model increased in radius by 
about 0.07\% per year and our 25~\msun\ model increased in radius by about 0.3\% 
per year. 

\subsection{Accounting for the Rotation}
\label{account}

There are several possible explanations for these results. One is that
we have not treated the physics of the stellar evolution properly. For
instance, there may be more viscosity than computed in the default
prescriptions in $MESA$ that would allow a greater transport of angular
momentum from the rotating core to the envelope. We have checked 
this aspect by computing the angular momentum in the cores of the
models at the stage of carbon burning. For a typical model, 20~\msun, 
the angular momentum of the helium core of mass 6.2~\msun\ with a 
surface velocity of 4.0~\kms\ is $\approx 1.8\times10^{50}$ g cm$^{-2}$ 
s$^{-1}$ and that of the carbon core of mass 2.3~\msun\ and surface velocity
of 7.0~\kms\ is $\approx 5.0\times10^{48}$ g cm$^{-2}$ s$^{-1}$. In carbon 
burning, the 20~\msun\ model has a total mass of 16.6~\msun\ and a hydrogen 
envelope of 10.4~\msun\ with an outer radius of $7\times 10^{13}$ cm.  
If the envelope had the same total angular momentum as the core, it would 
rotate at only $\approx 8\times10^{-4}$ the speed, or about 0.003~\kms,
even smaller than predicted by the models. Even if all the angular 
momentum of the helium core core were transferred to the extended envelope, 
the moment of inertia of the former is too small and of the latter is 
too large. The envelope would still not rotate substantially faster than 
$\sim0.1$~\kms. Massive stars are observed to rotate as rapidly as 
500~\kms\ near the ZAMS \citep{2013A&A...550A.109D}, but while we have 
not explored these extreme initial values, it is unlikely that an increase 
of the initial rotation velocity by a factor of a few will lead to changes 
in the final envelope rotation velocity by the required factor of $\sim 150$.

Another possibility is that the observed velocity is incorrect. The
velocity was determined by \citet{1996ApJ...463L..29G} and
\citet{1998AJ....116.2501U} by the use of {\sl HST} long--slit spectroscopy 
to map across the (minimally) resolved surface of Betelgeuse. The result was 
a systematic blueshift on one limb and a redshift on the opposite of the 
quoted magnitude. It is at least within the bounds of credibility that the 
measurements were affected, perhaps even dominated, by the large--scale 
convective motions of the envelope for which Betelgeuse is famous.

A particularly interesting prospect is to invoke the possibility
of duplicity. It is now known that a majority of O and B stars
are born in binary systems \citep{2012Sci...337..444S,
2015A&A...580A..93D}. \citet{2014ApJ...782....7D} estimate that
19\% of massive, apparently single stars (those with radial velocity
less than 10~\kms) are merged stars. \citet{2015ApJ...807L..21C}
report 17 giant stars observed with {\sl Kepler} that show rapid 
rotation that might signal coalescence in a binary system. 

We checked this possibility for Betelgeuse by estimating the mass of 
a putative companion star that, having coalesced with Betelgeuse from 
an orbit close to the current radius, would have provided the requisite 
angular momentum to spin the envelope up to $v_{rot}\sim 15$ \kms. If we 
assume the companion of mass $M_2$ is a point mass and thus neglect any 
internal angular momentum, that the envelope of Betelgeuse was originally 
non--rotating, and that all the orbital angular momentum of the companion
is deposited in the envelope, we can write
\begin{equation}
I_2 \Omega_2 = I_2\frac{v_{orb}}{R_{env}} = I_{env} \Omega_{env} = I_{env}\frac{v_{rot}}{R_{env}}
\end{equation}
or, with $I_2 = M_2 R_{env}^2$, $I_{env} \approx \frac{2}{5}M_{env}R_{env}^2$
and $v_{orb} = \sqrt{G M_{tot}/R_{env}}$,
\begin{equation}
M_2 \approx 1.2~M_\odot~M_{env,10} v_{rot,15} M_{tot,20}^{-1/2} R_{env,14}^{1/2}
\end{equation}
where $M_{env,10}$ is the envelope mass in units of 10~\msun, $M_{tot,20}$ is
the total mass of the system in units of 20~\msun, $v_{rot,15}$ is the final rotational 
velocity of the envelope in units of 15~\kms, and $R_{env,14}$ is the radius of the 
primary and of the orbit of the secondary at the epoch of coalescence. This 
simple model suggests that the current rotation of Betelgeuse could be 
explained if Betelgeuse were born with a companion of mass $\sim 1$~\msun\ 
with which it merged as it evolved up the RSB to its current position of glory.

While this hypothesis is credible and consistent with the {\sl a priori}
estimate that Betelgeuse has a probability of $\sim 20$\% of being 
born in a binary system, it raises a number of interesting issues. It
is not clear that that mass and angular momentum of a companion would
remain in the envelope. A companion of about a solar mass would have a
mean density of about 1 \gcm3. That density is characteristic of the base
of the hydrogen envelope in the models we consider here, implying that
a swallowed companion might not be dissolved until it reached the edge of
the helium core. If the companion plunged down to the core, the evolution 
might be severely altered by anomalous burning and mixing effects. The 
luminosity of an evolved massive star is typically a strong function
of the mass of the helium core and not the mass of the envelope. If a
companion partially or totally merged with the core of Betelgeuse, then
the current luminosity may be a measure of the core mass ($\sim$ 5 to 6~\msun),
but the mass of the envelope would be rather unconstrained and probably
smaller than the estimates given here based on basic, single--star models
that attempt to reproduce the luminosity, radius and effective temperature. 
Guessing the prior and future evolution of Betelgeuse becomes more problematic.
We also note that a coalescence might have affected surface abundances, 
complicating their interpretation.

If there were a coalescence, there might well be some mass ejected. If a 
shell were ejected with a velocity near the escape velocity, $v_{esc} \sim 
10 v_{10}$~\kms\ about $10^5 t_5$ years ago, then if that shell 
expanded unimpeded it would now be at a radius, $R_{sh} = 3\times10^{18}
v_{10} t_5$ cm. At a distance of 197 pc, the shell would have an angular
radius of about 17 arc minutes. If the shell swept up matter, then this
would be a rough upper limit to its location.

There is a well--observed ``bow shock" about 7 arc minutes away from Betelgeuse 
\citep{1997AJ....114..837N,2012A&A...548A.113D}. This bow shock is attributed to 
the wind from Betelgeuse sweeping up matter from the ISM in the direction of motion 
\citep{2012A&A...541A...1M,2012A&A...548A.113D,2014Natur.512..282M}, but it is 
possible that the flow of mass was impulsive rather than steady. The bow shock 
is rather smooth compared to numerical predictions of dynamic instabilities 
and is rather close to being circular \citep{2008PASJ...60S.407U}, again in 
contrast to wind simulations. It is not clear that an impulsive mass ejection 
at the time of companion merging would solve these issues, but this possibility 
(that is beyond the scope of the current paper) is worth exploring. There is 
also an odd, very linear feature beyond the bow shock that remains unexplained 
\citep{1997AJ....114..837N,2012A&A...548A.113D}. The observations also show a 
smaller ring of material at a radius of about 4 arc minutes \citep{2012MNRAS.422.3433L}. 
One explanation is that this is wind mass that is radiation--impeded by external 
radiation \citep{2014Natur.512..282M}. Such a structure could also form in 
the wind that was blown subsequent to the impulsive ejection we contemplate. Yet 
another possibility is that this inner ring is associated with the mass loss 
at the epoch of coalescence. This would require that the mass was ejected 
only about $2.4\times10^4$ years ago, or that there was considerable
deceleration of the ejected shell.

\section{Discussion}

We have used $MESA$ evolutionary models to explore the mass and evolutionary 
state of Betelgeuse. While $T_{eff}$, $R$, and $L$ are reasonably easy to 
reproduce for a range of masses for judicious choices of the still rather 
uncertain distance, we found the observed rotational velocity difficult to 
reproduce. Basic rotating models yielded the observed value, $v_{rot} \sim 15$~\kms\
only near the base of the giant branch. This position in the Hertzsprung--Russell
diagram, while unexpected, could not be completely ruled out given generous 
estimates of the uncertainty in distance. The principal objection to this result
is that the models are only in the range of the observed value of the rotation
velocity for a very brief time, a few hundred years. This is as improbable as
the likelihood that Betelgeuse is within a few hundred years of explosion.

A major uncertainty in this analysis is the distance to Betelgeuse. The 
estimate and associated uncertainties employed here were based on the results 
of observations with the {\sl VLA} and {\sl Hipparchos}. Unfortunately, 
Betelgeuse is brighter than the saturation limit of {\sl Gaia} so that mission may not
bring any improvement. Other more indirect methods to improve distance 
estimates would be most welcome.

We considered various factors that might mitigate our conclusion, including
uncertainties in the evolutionary models and in the observations that determine
the rotational velocity. The attempt to measure the observed rotation velocity 
may have been affected, or even dominated, by the convective motions of the 
envelope. Examination of unpublished STIS data tends to confirm the
original interpretation of rotation (A. Dupree, private communication, 2016).
This issue would be clarified by repeated observations to determine 
if the derived velocity is stable in time. The apparent rapid rotation velocity 
might also be affected by the presence of a dusty disk around the star.

We explore a solution based on the hypothesis that Betelgeuse was once a 
component of a binary star system and show that if it merged with a companion 
of about 1~\msun\ when it became a red supergiant that the requisite angular 
momentum of the envelope might be attained. We discuss the possibility that 
the various shells surrounding Betelgeuse might have been associated with this 
merging process. Betelgeuse might have been born in a dense cluster, an 
environment that could enhance the formation of multiples \citep{2008AJ....135.1430H}. 

If Betelgeuse has coalesced with a former binary companion, the task, while 
perhaps even more important, becomes yet more complicated. An appropriate 
quantitative analysis might enable the determination of the current state 
of evolution and hence the time to explosion, the original goal. In a future
paper, we will explore the promise of asteroseismology to illuminate this issue.

\section*{Acknowledgments}

We are grateful to E. L. (Rob) Robinson for insights, to Pawan Kumar for
discussions of stellar oscillations, to Fritz Benedict for a discussion of 
errors in distance measurements, and to Brian Mulligan
for help with Bibtex. We are especially thankful for the ample support
of Bill Paxton and the MESA team. This research was supported in
part by NSF Grant AST-11-9801.




\end{document}